\documentclass[a4paper]{jpconf}
\usepackage{graphicx}
\begin{document}
\title{A fluid of black holes at the beginning of the Universe}
\author{Pablo D\'\i az, Miguel Angel Per and Antonio Segu\'\i }

\address{
Departamento de F\'\i sica Te\'orica
Universidad de Zaragoza. 50009-Zaragoza. Spain
}

\ead{segui@unizar.es}

\begin{abstract}
The most entropic fluid can be related to a \emph{dense} gas of black holes that we use to study the beginning of the universe. We encounter difficulties to compatibilize an adiabatic expansion with the growing area for the coalescence of black holes. This problem may be circumvented for a quantum black hole fluid, whose classical counterpart can be described by a percolating process at the critical point. This classical regime might be related to the energy content of the current universe. 
\end{abstract}

\section{Introduction}

To understand the initial conditions of the universe we need a description of the gravitational field when the curvature is the same order or smaller than the Planck length. The most important candidate for a consistent description of quantum gravity is the string theory where the fundamental particles are no longer point-like but have a linear extension. There is a small number of consistent ways to quantize strings, all related by dualities, signaling an underlying structure that has been termed M theory. Despite this strong restriction in the consistent way to quantize the gravitational field, the true vacuum in which the theory lives is not predicted. Instead, a great amount of vacua are possible all compatible with the premises of the theory. The situation is such, that arguments outside the theory, mainly anthropic, have been used to select a vacuum \cite{weinberg}. Our purpose is to address the initial conditions using entropic arguments instead.

One of the basis underlying the theory of quantum gravity is the holographic principle\cite{thooft}. This states that a quantum description of the gravitational phenomena can be mapped into a dual theory on the boundary. In the string scenario such dual description is exemplified by the  celebrated AdS/CFT correspondence \cite{maldacena}. If a theory can be mapped onto its boundary, the operative number of degrees of freedom grows with the area of such boundary, in contrast to the case of ordinary quantum fields which scales with the volume of the system. A way to understand this drastic reduction on the number of degrees of freedom is by the formation of black holes (BH's). Due to the fact that the information carriers have energy, when we try to accumulate an amount of information in a smaller volume the subsystem can collapse forming a BH and lose the corresponding information. An amount of entropy is generated in this process.

In the semiclassical regime, when the curvature is larger than the Planck length, the holographic principle translates into different types of entropy bounds all sharing a property: the limitation on the number of degrees of freedom by the area which circumventes the system; an improvement of the situation was given by Bousso's covariant entropy bound \cite{bousso} . 

A BH is the most entropic system for a given amount of energy \cite{thooft}; if information is not lost in the subsequent evaporation, the BH is the most efficient recipient of information. Such information would be encoded in its event horizon and ejected away in subtle, higher non local correlation of its atmosphere. A gas of BH's will be the more entropic fluid and such universe will be able to attain the maximum complexity \cite{banks}.

\section{The equation of state of the most entropic fluid as a dense BH fluid}
In FRW cosmological models, the energy density $\rho$ and pressure $p$ of the cosmic fluid can be related by $p=\omega \rho$ with $\omega$ constant. We can associate the equation of state of the stiffest fluid, $\omega=1$ (sound velocity equal to light velocity), with a dense gas of BH's, what would constitute the most entropic fluid. This will be a candidate for the initial state of the universe \cite{banks}.
 
The dense fluid of BH satisfy 
\begin{equation}
n \sim \frac {1}{R_s^{d-1}}, \label{1}
\end{equation}
where $n$ is the number density of BH, $R_s$ its typical Schwartzschild radius, and $d$ the number of space-time dimensions. Now we show that the relation between its entropy density $\sigma$ and its energy density is given by 
\begin{equation}
\sigma \sim \rho^{1/2}, \label{2}
\end{equation}
independent on the number of dimensions. The relation between mass and size is $R_s \sim M^{1/(d-3)}$ (Newton constant has dimensions of energy to the $2-M$); the  energy density will be $ \rho = n M \sim R_s^{-2}$; the entropy density is given by $\sigma=n S$, where $S$ is the typical entropy of a BH $S = \Omega_{d-2} R_s^{d-2} /4$ with $\Omega_{d-2}$ being the area of the $d-2$ dimensional unit sphere; by substitution we have $\sigma \sim  R_s^{-1}$ and consequently, (\ref{2}) is satisfied.

On the other hand we can show that (\ref{2}) is equivalent to the stiffest equation of state for a fluid $p=\rho$ if both, the first law of thermodynamics apply and the expansion is adiabatic. The first law $dE=T dS-p dV$ together with the extensivity relation $E(\lambda S,\lambda V)= \lambda E(S,V)$ allows us to write 
\begin{equation}
\rho= T \sigma - p. \label{3}
\end{equation}
If relation (\ref{2}) adopts the form $\sigma= \alpha \rho^{1/2}$, were $\alpha$ is a constant, and taking the densities as $\sigma= S/V$ and $\rho =E/V$ we have $E=S^2/\alpha^2 V$ and $T=2 \rho^{1/2} /\alpha$. Substituting the previous values of $T$ and $\sigma$ in (\ref{3}) we obtain the promised relation $p= \rho$.

However, the two perspectives for the fluid, as a the dense fluid of BH's and as the stiffest fluid ($\omega=1$), present difficulties to represent the same physical system. A fluid of BH's in a coalescence process, if made only of BH's, is incompatible with the conservation of \emph{both}, energy and entropy. The fluid with equation of state $p=\rho$ as the driver of a FRW cosmological scenario, unless other fluid is present, describes an \emph{adiabatic} expansion. In the next section we study different mechanisms to solve this problem.

\section{Three ways to address the contradiction}

The first possibility is to allow no conservation of energy as the universe expands. For time varying cosmologies there is not time invariance and the corresponding energy conservation law is absent. This is the case, for example, for a radiation dominated universe, where due to the gravitational redshift of the photons, the density of energy of a comoving volume is not constant; at the same time, the entropy of the same comoving volume, proportional to the number of photons which do not vary with the cosmic evolution, remains constant.

Another way to liberate the tension between the two previous descriptions is to allow for a second cosmic fluid, so that we describe a two component fluid cosmological model; one of the components will be the BH fluid and the other a reservoir of energy in such a way that the total energy is conserved. The amount of entropy of this second component is arbitrary but it is reasonable to consider its value very small compared with to the entropy of the BH's. In this case, if the fundamental process is the Hawking's process \cite{hawking}, that is, the coalescence of two BH's of equal mass, we have the following results. We consider as the basic process, symbolically
\begin{equation}
\textrm{BH}(M)+\textrm{BH}(M)=\textrm{BH}(M_f)+\textrm{Energy}(E_r), \label{4}
\end{equation}
where $M$ and $M_f$ are the initial and final BH mass, and $E_r$ is the energy radiated in the process of coalescence and is carried by the second fluid. The conservation of energy is given by the relation (we use Planck units)
\begin{equation}
2M=M_{f}+E_r. \label{5}
\end{equation}
Now we can also impose the adiabaticity of the process (\ref{4}); using the fact that the entropy of a BH equals one quarter its area measured in Planck units we have
\begin{equation}
2R^{2} = R_{f}^{2} + {S_{r} \over \pi}, \label{6}
\end{equation}
where $S_r$ is the entropy carried by the second fluid that we consider negligible. Using (\ref{5}) and (\ref{6}) we can determine how the total energy is partitioned between the two fluids:
\begin{equation}
\Omega_{BH}={M_{f} \over 2M}= {R_{f}/2 \over R}={ 1 \over \sqrt{2} }=0.707, \label{7}
\end{equation}
is the part of the energy carried by the BH fluid, and $\Omega_{r}= E_{r}/ 2M =  1 -1/ \sqrt{2} =0.293$ is the energy radiated away in the process.

Finally, let us consider an intrinsic violation of energy conservation due to quantum effects, so that we will refer to a one component \emph{quantum} BH fluid. This violation $\Delta E$ is  quantically allowed by the uncertainty relations, if the violation takes place for a temporal interval $\Delta t \sim 1 / \Delta E$; $\Delta t$ can be considered as the cosmic time since the big bang. It is interesting to speculate on the classical counterpart of this quantum BH fluid that could be described by a percolating system of BH's at its critical point; a percolating system at its critical point presents scale invariance which could be related to the actual distribution of inhomogeneities. The  percolating critical point presents also a dual description in which the regions covered by BH's (black regions) are equivalent to regions surrounded by BH's (white regions). This dual white regions are covered by an event horizon and can be related to the actual accelerated state of the universe; in this way the coincidence problem, the fact that the dark energy ($\Omega_{DE}$) is about the same order as the ordinary energy ($\Omega_m$), can be addressed with the values obtained previously, $\Omega_{BH}= 0.707$ and $\Omega_r = 0.293$ \cite{coincidencia}.

\ack This work has been partially supported by MCYT (Spain) under grant FPA2003-02948.

\end{document}